\documentclass[11pt,english]{article}
\usepackage{graphics}
\usepackage[T1]{fontenc}
\usepackage{babel}
\usepackage[latin1]{inputenc}
\usepackage{amssymb}
\usepackage{times}
\usepackage{mathptm}

\newcommand{\dagg}{{\scriptscriptstyle\dagger}}

\begin{document}

\title{Effective rate equations for the over-damped motion in fluctuating potentials}

\author{Andreas Mielke\\
 Institut für Theoretische Physik, Ruprecht Karls Universität\\
 Philosophenweg 19, D-69120 Heidelberg, F.R. Germany \\
~\\
Dedicated to Prof. Dr. Erwin Müller-Hartmann on the occasion of his 60\protect\( ^{\rm th}\protect \)
birthday}

\maketitle
\begin{abstract}
We discuss physical and mathematical aspects of the over-damped motion of a
Brownian particle in fluctuating potentials. It is shown that such a system
can be described quantitatively by fluctuating rates if the potential fluctuations
are slow compared to relaxation within the minima of the potential, and if the
position of the minima does not fluctuate. Effective rates can be calculated;
they describe the long-time dynamics of the system. Furthermore, we show the
existence of a stationary solution of the Fokker-Planck equation that describes
the motion within the fluctuating potential under some general conditions. We
also show that a stationary solution of the rate equations with fluctuating
rates exists. 
\end{abstract}

\section{Introduction}

Thermally activated relaxation processes are an important mechanism for the
dynamics of many physical, chemical, and biological systems. The diffusion of
a Brownian particle in a potential with several minima and barriers serves as
a paradigm for such relaxation processes. The minima of the potential represent
stable or metastable states of the system, and the motion of the particle models
the transition of the system from one state to the other and back. The most
simple example in this class of models is the problem of diffusion over a single
potential barrier, pioneered by Kramers \cite{Kramers}. The dynamics of the
diffusion over a barrier is dominated by a characteristic time scale, which
is given by the mean first passage time for the escape out of the minimum of
the potential. In a model with several minima and barriers, the corresponding
time scales are given by the inverse rates for the transition from one state
to another. 

Often one uses kinetic rate equations to describe the dynamical properties of
relaxation in chemical or biological systems. This description is clearly much
simpler than the description by a Brownian particle in a potential. To calculate
rate coefficients, it is often sufficient to know some of the properties of
the potential energy surface, like barrier heigths. In principle, if one knows
the potential energy surface, both descriptions are equivalent. One can calculate
static or dynamic properties of the system either using the Brownian particle
in the potential or using the rate equations.

In many situations, the potential fluctuates due to some external fluctuations,
chemical reactions, or oscillations. These fluctuations usually have a finite
correlation time, which corresponds to a non-thermal noise. The most simple
model with a fluctuating potential consists of a single, fluctuating barrier,
where the height of the barrier fluctuates between two different values. Doering
and Gadoua \cite{DoeringGadoua} investigated such a simple model. They found
a local minimum in the mean first passage time as a function of the barrier
fluctuation rate. This effect has been called resonant activation and has been
studied extensively \cite{Zuercher93}-\cite{Marchi96}.

In principle it should be possible to use a description with rate equations
in the case with fluctuating barriers too. This picture has already been suggested
by Bier and Astumian \cite{Bier93}. They showed that the long-time dynamics
of a simple model with a dichotomously fluctuating linear ramp can be described
by kinetic rate equations if the potential fluctuations are not too fast. A
similar observation was made in \cite{Mielke99}, where it was shown that noise
induced stability of a metastable state occurs in systems with a single fluctuating
barrier.

In the present paper we consider potentials in \( d \) dimensions with several
minima and barriers. Furthermore, we allow general potential fluctuations. For
simplicity we assume that the potential fluctuations can be parametrized by
a single Markovian noise process \( z(t) \). Often it is argued that the Markovian
noise \( z(t) \) represents the cumulative effects of many weakly coupled environmental
degrees of freedom. In that case the central limit theorem can be applied and
\( z(t) \) becomes an Ornstein-Uhlenbeck process. On the other hand, there
are many realistic situations where the fluctuation of the potential is triggered
by a single (or few) environmental degree of freedom. The most simple case of
a non-Gaussian fluctuating potential is a dichotomously fluctuating potential
as discussed e.g. in \cite{DoeringGadoua}. But in many applications, the potential
fluctuates not only between two different values. Since all these different
cases are of interest, we will investigate general noise processes. Furthermore,
it is clear that there are always many possibilities to parametrize a fluctuating
potential by a Markovian noise process. This is another reason why it is useful
to consider general noise processes.

A main purpose of the present paper is to show under which conditions it is
possible to pass from a model with a fluctuating potential to a set of rate
equations with fluctuating rates. We will discuss general models in arbitrary
dimensions and with potentials having several minima. The general picture we
support by our calculations is that rate equations can be used if the potential
fluctuations are slow compared to the relaxation within in minima of the potential
and if the positions of the minima of the potential do not fluctuate. The long-time
behaviour of the model with a fluctuating potential is then similar to that
obtained using rate equations. The fluctuation of the potential is mainly a
fluctuation of the barrier heights. The potential fluctuations may be fast compared
to relaxation across the barriers of the potential.

After these remarks it is clear but important to mention that the applicability
of rate equations depends also on the quantities one in interested in. As long
as one is interested in stationary or quasi-stationary properties, rate equations
yield good results. If one is interested in properties on time scales of the
order of the relaxation within the minima or smaller, rate equations cannot
be applied. 

Furthermore we analyse systems with fluctuating rates in detail and derive effective
rates (which do not fluctuate). The effective rates yield the relevant time
scales for the long-time behaviour of the system. They can be directly compared
to the mean escape rates in the fluctuating potentials. Furthermore they yield
the stationary distribution, which can also be compared with results obtained
directly by solving the Fokker-Planck equation with a fluctuating potential.
The comparison is easily done for simple, one-dimensional systems with piecewise
linear potentials. For such systems it is relatively easy to solve the Fokker-Planck
equation directly.

The paper is organized as follows: In the next section we discuss how a kinetic
rate equation with fluctuating rates can be obtained starting from a Fokker-Planck
equation with a fluctuating potential and in which situations one can expect
it to be valid. We also show how one can derive effective rates from the kinetic
rate equation with fluctuating rates. The effective rates yield the characteristic
time scales for the escape out of the minima of the fluctuating potential. The
general formula for the effective rates can be evaluated for special noise processes.
This is done in section 3. The results are used to compare the description of
a system by effective rates with the original Fokker-Planck equation. For our
discussion the existence of a stationary solution of the Fokker-Planck equation
with a fluctuating potential and the existence of a stationary solution of a
kinetic rate equation with fluctuating rates is essential. Section 3 also contains
the exact solution of the dichotomous two-state system and its derivation. Some
of the exact results have already been mentioned without derivation in \cite{Mielke99}.
In section 4 we show that the stationary solutions of the rate equation with
fluctuating rates and of the Fokker-Planck equation with a fluctuating potential
exists under some general assumptions on the potential. This result is a generalization
of a recent, similar result by Pechukas and Ankerhold \cite{Pechukas98}. Finally,
in section 5, some possible applications are mentioned and conclusions of our
results are drawn. As a specific example we discuss how our results can be used
to interprete experimental findings for membrane proteins.

\section{Derivation of effective rate equations}

\subsection{Noise processes and the Fokker-Planck equation}

The over-damped motion of a particle in a fluctuating potential is usually described
by a Langevin equation

\begin{equation}
\frac{d\vec{x}}{dt}=\vec{f}(\vec{x},t)+\sqrt{2T}\vec{\xi }(t).
\end{equation}
 We have chosen the units such that the friction constant is unity. \( \xi (t) \)
is a thermal (Gaussian white) noise, it satisfies \( \langle \vec{\xi }(t)\rangle =0 \),
\( \langle \xi _{a}(t)\xi _{b}(t')\rangle =\delta _{ab}\delta (t-t') \). \( \vec{f}(\vec{x},t)=-\nabla V(\vec{x},t) \)
is the force of the fluctuating potential. We investigate situations where the
potential tends to infinity for \( |\vec{x}|\rightarrow \infty  \) or where
the potential is defined in a finite domain \( \Omega  \) with reflecting boundary
conditions on the boundary \( \partial \Omega  \). Let us assume that the potential
fluctuations can be parameterized by a single stochastic variable \( z(t) \),
\( V(\vec{x},t)=V(\vec{x},z(t)) \). It is clear that such a representation
is not unique. Instead of \( z(t) \) one can use any monotonic function of
\( z(t) \) to parametrize \( V(x,t) \). Often one uses a simple linear ansatz
\( V(x,z(t))=V_{0}(x)+z(t)\Delta V(x) \), with a dichotomous or an Ornstein-Uhlenbeck
process \( z(t) \). As already mentioned in the introduction, we want to investigate
models for which the potential fluctuations represent the effects of few environmental
degrees of freedom. Therefore we dot not restrict ourselves to a linear ansatz
or to special noise processes. The only assumption we make is that the noise
process \( z(t) \) can be described using a Fokker-Planck equation for the
probability distribution \( p(z,t) \), 
\begin{equation}
\frac{\partial p(z,t)}{\partial t}=M_{z}p(z,t).
\end{equation}
 The right eigenfunctions of the generator \( M_{z} \) are denoted by \( \phi _{n}(z), \)
the eigenvalues by \( -\lambda _{n} \)
\begin{equation}
M_{z}\phi _{n}(z)=-\lambda _{n}\phi _{n}(z),
\end{equation}
 where \( \lambda _{0}=0 \), \( \lambda _{1}>0 \), \( \lambda _{n}\geq \lambda _{n-1} \).
\( \phi _{0}(z) \) is the stationary distribution of the noise process \( z(t) \).
We let \( \phi _{n}(z)=g_{n}(z)\phi _{0}(z) \). Then one has 
\begin{equation}
\label{right_eigenfunctions_M_z}
\int dzg_{n}(z)g_{m}(z)\phi _{0}(z)=\delta _{n,m}.
\end{equation}
 \( g_{n}(z) \) are orthogonal functions with respect to the weight function
\( \phi _{0}(z) \). For \( n=m \) this equation fixes the normalization of
\( \phi _{n}(z) \). The eigenvalue \( \lambda _{1}=\tau ^{-1} \) determines
the correlation time of the noise process \( z(t) \). We already mentioned
that the parametrization of the fluctuating potential by a stochastic variable
\( z(t) \) is not unique. One can always choose a different parametrization.
The final results should of course be independent of the parametrization. If
one changes the parametrization, the spectrum of \( M_{z} \) remains the same,
but the right eigenfunctions are changed. 

The motion of the over-damped particle in the fluctuating potential can be described
using the joint probability distribution \( \rho (\vec{x},z,t) \) for \( \vec{x}(t) \)
and \( z(t) \). It obeys the Fokker-Planck equation 
\begin{equation}
\label{eq:FokkerPlanck}
\frac{\partial \rho (\vec{x},z,t)}{\partial t}=\nabla \cdot \left( (\nabla V(\vec{x},z))+T\nabla \right) \rho (\vec{x},z,t)+M_{z}\rho (\vec{x},z,t).
\end{equation}
 \( V(\vec{x},z) \) must have the same properties as a function of \( \vec{x} \)
as \( V(\vec{x},t). \) It must tend to infinity sufficiently fast when \( |\vec{x}| \)
becomes large; or alternatively, we may define the Fokker-Planck equation on
a finite, open domain \( \Omega  \) with reflecting boundary conditions on
the boundary \( \partial \Omega  \). Furthermore we assume that \( V(\vec{x},z) \)
is finite for any \( z \) out of the support of \( \phi _{0}(z) \). \( V(\vec{x},z) \)
does not have any infinitely high potential wells inside \( \Omega  \).

\subsection{Rate equations }

\subsubsection{Rate equations for fixed potentials}

Let us first briefly review the case of a potential that does not fluctuate.
The Fokker-Planck operator has the form
\begin{equation}
L=\nabla \cdot \left( (\nabla V(\vec{x}))+T\nabla \right) .
\end{equation}
Let us assume that the potential \( V(\vec{x}) \) has \( N \) distinct minima
at the points \( \vec{x}_{i} \), \( i=0,\ldots ,N-1 \) and that \( V(\vec{x})\rightarrow \infty  \)
for \( |\vec{x}|\rightarrow \infty  \) sufficiently fast. \( L \) has one
eigenvalue \( 0 \), the corresponding right eigenfunction is the stationary
distribution \( \propto \exp (-V(\vec{x})/T) \). \( -L \) is non-negative,
i.e. all the other eigenvalues of \( -L \) are positive. Using a simple variational
argument, one can show that \( -L \) has at least \( (N-1) \) eigenvalues
which behave like \( \exp (-c/T) \) for small \( T \). Let us divide the entire
space \( \Omega  \) on which the system is defined into subspaces \( \Omega _{i} \).
\( \vec{x}_{i}\in \Omega _{j} \) if and only if \( i=j \). \( \Omega _{i}\cap \Omega _{j}=\emptyset  \)
if \( i\neq j \). \( \bigcup _{i}\Omega _{i}=\Omega \setminus \left( \bigcup _{i}\partial \Omega _{i}\right)  \).
\( \partial \Omega _{i} \) denotes the boundary of \( \Omega _{i} \). The
boundaries are chosen such that \( \vec{n}_{i}(\vec{x})\cdot \nabla V(\vec{x})=0 \)
for \( \vec{x}\in \partial \Omega _{i} \), where \( \vec{n}_{i}(\vec{x}) \)
is the normal vector to the boundary \( \partial \Omega _{i} \) at \( \vec{x} \).
Such a division of \( \Omega  \) is always possible. Depending on the form
of \( V(\vec{x}) \) it may not be unique. Let us now introduce subsets \( \tilde{\Omega }_{i}\subset \Omega _{i} \)
which do not contain a small region close to the boundary of \( \Omega _{i} \),
but which do contain the positions \( \vec{x}_{i} \) of the minima together
with a sufficiently large region around them. Let \( \tilde{\theta }_{i}(\vec{x}) \)
be a smooth function that is equal to unity if \( \vec{x}\in \tilde{\Omega }_{i} \)
and that vanishes if \( \vec{x}\notin \Omega _{i} \). In \( \Omega _{i}\setminus \tilde{\Omega }_{i} \)
the function \( \tilde{\theta }_{i}(\vec{x}) \) drops smoothly from \( 1 \)
to \( 0 \). We assume that sufficiently many derivatives of \( \tilde{\theta }_{i} \)
exist. 

In a next step, we introduce \( \hat{L}=\exp (V(\vec{x})/(2T)L\exp (-V(\vec{x})/(2T)) \).
\( \hat{L} \) is a hermitian operator. It has the explicit form
\begin{equation}
\hat{L}=T\nabla ^{2}+\frac{1}{2}(\nabla ^{2}V)-\frac{1}{4T}(\nabla V)^{2}.
\end{equation}
 Clearly, \( \hat{L} \) has the same eigenvalues as \( L \). \( \exp (-V(\vec{x})/(2T)) \)
is the (unnormalized) eigenfunction of \( \hat{L} \) to eigenvalue \( 0 \).
Let \( \hat{g}_{i}(\vec{x})=C_{i}\exp (-V(\vec{x})/(2T))\tilde{\theta }_{i}(\vec{x}) \),
where \( C_{i} \) is chosen such that \( \int d^{d}x|\hat{g}_{i}(\vec{x})|^{2}=1 \).
One has \( \int d^{d}x\hat{g}_{i}(\vec{x})\hat{g}_{j}(\vec{x})=\delta _{i,j} \).
Let \( \hat{L}_{i,j}=\int d^{d}x\hat{g}_{i}(\vec{x})\hat{L}\hat{g}_{j}(\vec{x}) \).
These quantities vanish for \( i\neq j \). For \( i=j \) one obtains 
\begin{equation}
\hat{L}_{ii}=\frac{T}{2}|C_{i}|^{2}\int d^{d}x\nabla \cdot (\exp (-V(\vec{x})/T)\nabla \tilde{\theta }_{i}(\vec{x})).
\end{equation}
Since \( \nabla \tilde{\theta }_{i} \) vanishes for \( \vec{x}\in \tilde{\Omega }_{i} \),
\( \hat{L}_{ii} \) contains a factor \( \exp (-\Delta \hat{V}_{i}/T) \) where
\begin{equation}
\Delta \hat{V}_{i}=\min _{x\in \Omega _{i}\setminus \tilde{\Omega }_{i}}V(\vec{x})-V(\vec{x}_{i}).
\end{equation}
Let \( \Delta \hat{V}=\min _{i}\Delta \hat{V}_{i} \). Then this shows that
\( -\hat{L} \) has \( N \) eigenvalues that decay at least as fast as \( \exp (-\Delta \hat{V}/T) \)
to zero when \( T \) tends to zero. 

Let us now consider \( L \) restricted to \( \Omega _{i} \) but with reflecting
boundary conditions on \( \partial \Omega _{i} \). With these boundary conditions
\( -L \) has one eigenvalue \( 0 \) and typically other eigenvalues which
are much larger than \( \exp (-\Delta \hat{V}/T) \). These eigenvalues describe
the time-scale for a decay within the potential well around the minimum at \( \vec{x}_{i} \).
On the original domain \( \Omega  \) one can therefore expect that \( -L \)
has one eigenvalue \( 0 \) and \( N-1 \) eigenvalues which are of the order
\( \exp (-c/T) \) and that all other eigenvalues of \( -L \) are much larger
if \( T \) is sufficiently small. Let us denote the smallest \( N \) eigenvalues
of \( -L \) by \( \lambda _{i} \), \( i=0,\ldots ,N-1 \), \( \lambda _{0}=0 \)
and the corresponding right eigenfunctions by \( g_{i}(\vec{x}) \). Then, for
time scales large compared to the intra-well relaxation times of the potential
\( V(\vec{x}) \), we can make the ansatz
\begin{equation}
\rho (\vec{x},t)=\sum _{i=0}^{N-1}\rho _{i}(t)g_{i}(\vec{x})
\end{equation}
 for the solution of the Fokker-Planck equation \( \frac{\partial \rho }{\partial t}=L\rho  \).
This ansatz should describe the long-time behaviour of the system quite well.
It is valid on time scales being large compared to the intra-well relaxation
times in the different potential wells of the potential. \( n_{i}(t)=\int _{x\in \Omega _{i}}d^{d}x\rho (\vec{x},t) \)
is the probability to find the particle in the region \( \Omega _{i} \), i.e.
close to the minimum \( \vec{x}_{i} \). Let \( n_{ij}=\int _{x\in \Omega _{i}}d^{d}xg_{j}(\vec{x}) \).
One has \( n_{i}(t)=\sum _{j}\rho _{j}(t)n_{ij} \). For the derivative with
respect to time one obtains
\begin{eqnarray}
\frac{dn_{i}}{dt} & = & \int _{x\in \Omega _{i}}d^{d}xL\rho \, =\, \sum _{j}\rho _{j}\int _{x\in \Omega _{i}}d^{d}xLg_{j}\nonumber \\
 & = & -\sum _{j}\lambda _{j}\rho _{j}n_{ij}\, =\, \sum _{k}r_{ik}n_{k},
\end{eqnarray}
 where the matrix \( R=(r_{ij})_{i,j=0,\ldots ,N-1} \) is given by \( R=-N^{-1}\Lambda N \).
\( N \) is the matrix \( (n_{ij})_{i,j=0,\ldots ,N-1} \) and \( \Lambda  \)
is a diagonal matrix with the entries \( \lambda _{j} \), \( j=0,\ldots ,N-1 \).
The differential equation \( \frac{dn_{i}}{dt}=\sum _{k}r_{ik}n_{k} \) is the
rate equation for this problem. By construction, the eigenvalues of \( R \)
are \( -\lambda _{i} \). 

Unfortunately, the definition of \( R \) is not useful for an explicit calculation.
If one is able to solve the eigenvalue-problem of \( L \), there is no need
to describe the problem by a rate equation. The rate equation is useful in situations,
where it is not possible to obtain analytic results from the Fokker-Planck equation.
Fortunately, approximative values for the rates \( r_{ij} \) often yield very
good results. In the one-dimensional case the usual Kramers' rate is sufficiently
good.

By construction, \( R \) is equivalent to the Fokker-Planck operator \( L \)
projected onto the space of its eigenfunctions with \( N \) largest eigenvalues.
This means that dynamical properties on time scales corresponding to eigenvalues
or larger can be well described by \( R \). Properties on time scales \( |\lambda _{N}|^{-1} \)
or smaller cannot be calculated using \( R \). \( \lambda _{N} \) sets the
time scale for relaxation within the potential wells.

\subsubsection{Rate equations for fluctuating potentials}

Under which conditions is it possible to describe a problem with a fluctuating
potential by rate equations? It is clear that one must be able to define \( n_{i} \)
for a fluctuating potential. This is possible if the stationary distribution
is strongly peaked at the minima \( \vec{x}_{i} \). But in general one can
have a situation, where the positions \( \vec{x}_{i} \) of the minima fluctuate
as well. If this is the case, the probability distribution will not necessarily
be peaked. Let us discuss that using a simple example: A one dimensional dichotomously
fluctuating potential.

For this example, \( z \) takes two values, \( z=\pm 1 \) with equal probability.
\( M_{z} \) has the eigenvalues \( 0 \) and \( -\tau ^{-1} \). In the eigenbasis
of \( M_{z} \) the potential \( V(x,z) \) has the form 
\begin{equation}
\left( \begin{array}{cc}
\left\langle V\right\rangle (x) & \Delta V(x)\\
\Delta V(x) & \left\langle V\right\rangle (x)
\end{array}\right) ,
\end{equation}
 where \( \left\langle V\right\rangle (x)=(V(x,+)+V(x,-))/2 \), \( \Delta V(x)=(V(x,+)-V(x,-))/2 \).
Let \( f(x)=-\frac{d\left\langle V\right\rangle (x)}{dx} \), \( \Delta f(x)=-\frac{d\Delta V(x)}{dx} \).
Let us calculate the stationary probability \( p_{0}(x)=\int dz\, \rho (x,z) \).
The stationary joint probability distribution can be written as \( \rho (x,z)=p_{0}(x)\phi _{0}(z)+p_{1}(x)\phi _{1}(z) \).
Inserting this into the stationary Fokker-Planck equation yields 
\begin{equation}
\frac{d}{dx}(f+z\Delta f-T\frac{d}{dx})(p_{0}(x)\phi _{0}(z)+p_{1}(x)\phi _{1}(z))+\tau ^{-1}p_{1}(x)\phi _{1}(z)=0.
\end{equation}
 This yields two equations for \( p_{0}(x) \) and \( p_{1}(x) \) corresponding
to the two coefficients of \( \phi _{0}(z) \) and \( \phi _{1}(z) \), which
both have to vanish. Eliminating \( p_{1}(x) \) yields a single equation for
\( p_{0}(x), \) which can be written as
\begin{equation}
\left[ 1+\tau \frac{d}{dx}\left( f-T\frac{d}{dx}\right) \right] \left( \Delta f\right) ^{-1}\left( f-T\frac{d}{dx}\right) p_{0}(x)+\tau \frac{d}{dx}\Delta f\, p_{0}(x)=0.
\end{equation}
 This equation can be derived directly from the Fokker-Planck equation. For
\( T=0 \) it can be solved explicitely. One obtains 
\begin{equation}
p_{0}(x)=C\frac{\Delta f}{f^{2}-\Delta f^{2}}\exp \left( -\frac{1}{\tau }\int dx\frac{f}{f^{2}-\Delta f^{2}}\right) .
\end{equation}
 The integrand \( f/(f^{2}-\Delta f^{2}) \) diverges at the minima of \( V(x,\pm ) \).
If the position of the minima does not change, \( p_{0}(\vec{x}) \) is a sum
of \( \delta  \)-function located at the minima with appropriate weights. With
changing positions of the minima, the situation becomes different. In that case
the particle moves between the minima. The most simple example to illustrate
that is \( V(x,\pm )=\frac{1}{2}\gamma _{\pm }(x-x_{\pm })^{2} \), \( \gamma _{\pm }>0 \).
Let \( x_{+}>x_{-} \). Then the support of \( p_{0}(x) \) is the interval
\( [x_{-},x_{+}] \). For \( p_{0}(x) \) one obtains 
\begin{equation}
p_{0}(x)=C\left( \frac{1}{\gamma _{+}(x_{+}-x)}+\frac{1}{\gamma _{-}(x-x_{-})}\right) (x_{+}-x)^{1/(2\tau \gamma _{+})}(x-x_{-})^{1/(2\tau \gamma _{-})}
\end{equation}
 with some normalization constant \( C \). The distribution function \( p_{0}(x) \)
is not peaked at \( x_{\pm } \). It is clear that for small \( T>0 \) one
obtains a similar result. This simple example shows that in a situation where
the position of the minima depends on \( z \) the stationary distribution of
the system is not necessarily peaked at the minima of the potential.

Let us now assume that the potential \( V(\vec{x},z) \) has minima at \( \vec{x}_{i} \)
and that the positions \( \vec{x}_{i} \) of the minima do not depend on \( z \).
For a fixed potential, rate equations can be used for time scales larger than
the intra-well relaxation times. It is therefore clear that for potential fluctuations
being faster than the intra-well relaxation times one cannot use rate equations.
If the potential fluctuations are slower than the intra-well relaxation times,
and if the position of the minima does not fluctuate, one is in an adiabatic
situation. The particle is able to follow the fluctuations of the potential
as long as it stays close to a minimum. This means that one can calculate the
rates for fixed \( z. \) The rates become functions of \( z \). The system
is thus described by rate equations with fluctuating rates,
\begin{equation}
\frac{dn_{i}}{dt}=\sum _{j}r_{ij}(z(t))n_{j}.
\end{equation}
 These rate equations are considerably simpler than the Fokker-Planck equation
with a fluctuating potential. On the other hand, the validity of these equations
is not entirely clear. We used only heuristic arguments to obtain the rate equations.
In the following section we will compare results obtained from the rate equations
with results from the Fokker-Planck equation. This allows us to show, in which
range of parameters the rate equations can be used.

\subsection{Elimination of \protect\( z(t)\protect \)}

If one is interested in stationary properties or in the long-time behaviour
of the system, it is useful to derive effective rates for the transition from
one state to another. The effective rates do not depend on \( z \). They can
be calculated by eliminating \( z(t) \). If the correlation time \( \tau  \)
of \( z(t) \) is small, the elimination process is a standard elimination of
a fast variable. To lowest order in \( \tau  \) the effective rates are simply
given by the average of the rates \( r_{ij}(z) \). To derive effective rates,
let us transform the rate equation 
\begin{equation}
\frac{d\vec{n}}{dt}=R\vec{n}
\end{equation}
 into a Fokker-Planck equation for the joint probability density \( p(\vec{n},z,t) \)
\begin{equation}
\frac{\partial p(\vec{n},z,t)}{\partial t}=\nabla _{\vec{n}}\cdot (R\vec{n}\, p(\vec{n},z,t))+M_{z}p(\vec{n},z,t).
\end{equation}
 Since \( \sum _{i}n_{i}=1 \) for all \( t \), \( p(\vec{n},z,t) \) contains
a factor \( \delta (\sum _{i}n_{i}-1) \). To calculate average values for \( \vec{n} \),
it is sufficient to consider the stationary case. Let 
\begin{equation}
p(\vec{n},z)=p_{0}(\vec{n})\phi _{0}(z)+\bar{p}(\vec{n},z)
\end{equation}
 where 
\begin{equation}
\int dz\, \bar{p}(\vec{n},z)=0,
\end{equation}

\begin{equation}
\int d^{N}n\, \bar{p}(\vec{n},z)=0.
\end{equation}
 Let \( \left\langle R\right\rangle =\int dz\, R(z)\phi _{0}(z) \). The stationary
Fokker-Planck equation yields 
\begin{equation}
\label{eq:p0}
\nabla _{\vec{n}}\cdot (\left\langle R\right\rangle \vec{n}\, p_{0}(\vec{n}))=-\nabla _{\vec{n}}\int dz\, R\vec{n}\, \bar{p}(\vec{n},z)
\end{equation}
 and 
\begin{equation}
\label{eq:pbar}
\bar{p}(\vec{n},z)=-M_{z}^{-1}\nabla _{\vec{n}}R\vec{n}\, (p_{0}(\vec{n})\phi _{0}(z)+\bar{p}(\vec{n},z)).
\end{equation}
 Here, \( M_{z}^{-1} \) is the generalized inverse, it obeys \( M_{z}^{-1}\phi _{n}(z)=-\lambda _{n}^{-1}\phi _{n}(z) \)
for \( n>0 \), \( M_{z}^{-1}\phi _{0}(z)=0 \). (\ref{eq:p0}) yields 
\begin{equation}
\label{eq:Rnp}
\left\langle R\right\rangle \vec{n}\, p_{0}(\vec{n})=-\int dz\, R\vec{n}\, \bar{p}(\vec{n},z)+\vec{j}(\vec{n})
\end{equation}
 with \( \nabla _{\vec{n}}\vec{j}(\vec{n})=0 \). Let \( \left\langle \vec{n}\right\rangle =\int d^{N}n\, \vec{n}p_{0}(\vec{n}) \).
Since 
\begin{equation}
\int d^{N}n\, \vec{j}(\vec{n})=\left\langle \frac{d\vec{n}}{dt}\right\rangle =0
\end{equation}
 one has 
\begin{eqnarray}
\left\langle R\right\rangle \left\langle \vec{n}\right\rangle  & = & \sum _{l=0}^{\infty }\int dz\int d^{N}n(-1)^{l}R\vec{n}(M_{z}^{-1}\nabla _{\vec{n}}R\vec{n})^{l+1}p_{0}(\vec{n})\phi _{0}(z)\nonumber \\
 & = & -\sum _{l=0}^{\infty }\int dz\int d^{N}n(RM_{z}^{-1})^{l+1}R\vec{n}p_{0}(\vec{n})\phi _{0}(z).\label{eq:Rnav} 
\end{eqnarray}
 The first line in (\ref{eq:Rnav}) is obtained by solving (\ref{eq:pbar})
iteratively and inserting the result in (\ref{eq:Rnp}). Integrating by parts
yields the second line. Since \( R \) and \( M_{z} \) do not depend on \( \vec{n} \),
the integration over \( \vec{n} \) yields the average \( \left\langle \vec{n}\right\rangle  \).
The sum over \( l \) gives finally 
\begin{equation}
\label{eq:avn}
\int dz\left( 1-RM_{z}^{-1}\right) ^{-1}R\phi _{0}(z)\left\langle \vec{n}\right\rangle =0.
\end{equation}
 The entries of the matrix 
\begin{equation}
\label{eq:effrates}
R_{\rm eff}=\int dz\left( 1-RM_{z}^{-1}\right) ^{-1}R\phi _{0}(z)
\end{equation}
 are the effective rates, the negative inverse of its diagonal matrix elements
are the mean escape times out of the states \( i \). The expansion in (\ref{eq:Rnav})
can be viewed as a usual \( \tau  \)-expansion. Expanding the right-hand side
of (\ref{eq:effrates}) yields the \( \tau  \)-expansion for the effective
rates. But since \( (1-RM_{z}^{-1}) \) is positive definite, the right hand
side of (\ref{eq:effrates}) is well defined for any \( \tau  \). Whereas it
is not possible to speak of an effective barrier for large \( \tau  \) \cite{Pechukas94},
it is possible to obtain effective rates. But, depending on the noise process
and on \( R(z) \), it cannot be excluded that the smallest non-vanishing eigenvalue
of \( R_{\rm eff} \) tends to zero in the limit \( \tau \rightarrow \infty  \).
This may happen if the fluctuations of the barrier heights in the potential
can be arbitrarily large, so that infinite barrier heights occur. In that case
the solution for \( \left\langle \vec{n}\right\rangle  \) has to be obtained
from the solution for finite \( \tau  \) and the limit \( \tau \rightarrow \infty  \)
has to be taken afterwards, since (\ref{eq:avn}) does not have a unique solution
for \( \tau =\infty  \). As long as the eigenvalue 0 of \( R(z) \) is non-degenerate
for all \( z \) in the support of \( \phi _{0}(z) \), the matrix \( R_{\rm eff} \)
has a non-degenerate eigenvalue 0 and (\ref{eq:avn}) has a unique solution.
This is shown below in section 4.

\subsection{Applicability of effective rate equations}

As we already discussed in subsection 2.1, rate equations describe only properties
on long time scales. If one wants to investigate dynamical properties on characteristic
time scales of the order of the typical relaxation within a potential well,
one cannot use rate equations. If one wants to study a system with a fluctuating
potential that fluctuates fast (i.e. one a time scale of the order of the typical
relaxation within a potential well) one has to use the Fokker-Planck equation.
This is true for equations with fluctuating rates as well as for the effective
rate equations we derived in the last subsection. 

Since in many applications details of the potential are unknown, one often uses
rate equations. Especially in biological applications, e.g. for the description
of membrane proteins, one usually uses rate equations. A typical example is
the well known paper by Petracchi et al \cite{Petracchi94}, who studied the
effect of time dependent electric fields on membrane proteins experimentally
and used rate equations to interprete their experimental findings. They were
able to explain most of their results using rate equations and a very simple
\emph{ansatz} for the rates. In their experimental studies they found an interesting
effect called phase anticipation. The rate equations on the other hand do not
show this effect. They argued that phase anticipation occurs due to some special
biological effect. From our discussion about the validity of rate equations
it is clear that the effect of phase anticipation, which occurs on short time
scales, cannot be described by rate equations. We will come back to this point
in our conclusions.

In the following section we discuss some simple, one-dimensional examples. The
most simple case, the dichotomous two-state model, is exactly solvable. For
this case and for more general examples we compare the solution of the rate
equations with the solution of the Fokker-Planck equation. For all cases, rate
equations yield very good results if the potential fluctuations are slow compared
to the relaxation within the potential wells.

\section{Solutions for special cases}

\subsection{The dichotomous two-state model }

The most simple model with a fluctuating potential is a dichotomous two-state
model. The potential \( V(\vec{x},z) \) has two minima at \( \vec{x}_{1} \)
and \( \vec{x}_{2} \). \( z(t) \) is a dichotomous process. Such a model has
two states, and corresponding rates \( r_{ij} \), \( i,j=1,2 \) for transitions
between them . For this model we let \( r_{1}:=r_{21}=-r_{11} \), \( r_{2}:=r_{12}=-r_{22} \).
The rate equations are

\begin{equation}
\frac{dn_{1}}{dt}=-r_{1}n_{1}+r_{2}n_{2},
\end{equation}

\begin{equation}
\frac{dn_{2}}{dt}=-r_{2}n_{2}+r_{1}n_{1}.
\end{equation}
 Using the normalization condition \( n_{1}+n_{2}=1 \) one obtains a single
differential equation 
\begin{equation}
\frac{dn_{1}}{dt}=-(r_{1}+r_{2})n_{1}+r_{2}.
\end{equation}
 In the following the index of \( n_{1} \) will be dropped. The Fokker-Planck
equation for \( p(n,z,t) \) is 
\begin{equation}
\frac{\partial p(n,z,t)}{\partial t}=\frac{\partial }{\partial n}((r_{1}+r_{2})n-r_{2})p(n,z,t)+M_{z}p(n,z,t).
\end{equation}
 In general it is possible to expand the stationary solution using the eigenfunctions
of \( M_{z} \)
\begin{equation}
\label{eq:stat_{a}nsatz}
p(n,z)=\sum _{k}p_{k}(n)\phi _{k}(z).
\end{equation}
 For the matrix elements of \( r_{i}(z) \) one has the representation 
\begin{equation}
\label{eq:matrixelements}
r_{i}(z)\phi _{k}(z)=\sum _{l}r_{ikl}\phi _{l}(z).
\end{equation}
 It is useful to define 
\begin{equation}
r_{kl}:=r_{1kl}+r_{2kl}.
\end{equation}
 The Fokker-Planck equation for the stationary distribution \( p(n,z) \) has
now the form 
\begin{eqnarray}
\frac{\partial }{\partial n}((r_{1}+r_{2})n-r_{2})p(n,z) & = & \frac{\partial }{\partial n}\sum _{kl}(r_{kl}n-r_{2kl})p_{k}\phi _{l}\nonumber \label{eq:general_{s}tat_{d}gl} \\
 & = & \sum _{l}\lambda _{l}p_{l}\phi _{l}.
\end{eqnarray}
 The two matrices \( R_{i}=(r_{ikl}) \) have the same eigenvectors and only
positive eigenvalues. This must be true for \( R=R_{1}+R_{2} \) as well. In
the dichotomous case \( R \) and \( R_{2} \) are \( 2\times 2 \)-matrices.
The eigenvalues are \( r_{\pm } \) and \( r_{2\pm } \). \( r_{i}(z) \) takes
the two values \( r_{i\pm } \) with the probabilities \( p_{\pm } \). Let
\( n_{\pm }=r_{2\pm }/r_{\pm } \). I assume \( n_{+}>n_{-} \). The explicite
form is of \( R \) is 
\begin{equation}
R=\left( \begin{array}{cc}
r_{00} & r_{01}\\
r_{10} & r_{11}
\end{array}\right) =\left( \begin{array}{cc}
p_{+}r_{+}+p_{-}r_{-} & \sqrt{p_{+}p_{-}}(r_{+}-r_{-})\\
\sqrt{p_{+}p_{-}}(r_{+}-r_{-}) & p_{+}r_{-}+p_{-}r_{+}
\end{array}\right) 
\end{equation}
 and similarly for \( R_{2} \). \( p_{0}(n) \) vanishes for \( n<n_{-} \)
and for \( n>n_{+} \). The two equations (\ref{eq:general_{s}tat_{d}gl}) are
\begin{equation}
\label{eq:p01first}
(r_{00}n-r_{200})p_{0}+(r_{10}n-r_{210})p_{1}=0,
\end{equation}
\begin{equation}
\frac{d}{dn}\left( (r_{01}n-r_{201})p_{0}+(r_{11}n-r_{211})p_{1}\right) =\lambda _{1}p_{1}.
\end{equation}
 The integration constant in (\ref{eq:p01first}) vanishes since the average
of \( \frac{dn}{dt} \) vanishes. It may be used to eliminate \( p_{1} \),
one obtains 
\begin{equation}
\frac{d}{dn}Ap_{0}=-\lambda _{1}Bp_{0}
\end{equation}
 where 
\begin{equation}
A=\frac{f(n)}{r_{10}n-r_{210}},
\end{equation}
\begin{equation}
B=\frac{r_{00}n-r_{200}}{r_{10}n-r_{210}}.
\end{equation}
 The solution of this equation is 
\begin{equation}
\label{eq:p0dicho_gen}
p_{0}=\frac{C}{A}\exp (-\lambda _{1}\int \frac{B}{A}dn)
\end{equation}
 where \( C \) is a normalization constant and \( f(n) \) is given by 
\begin{eqnarray}
f(n) & := & (r_{01}n-r_{201})(r_{10}n-r_{210})-(r_{00}n-r_{200})(r_{11}n-r_{211})\nonumber \\
 & = & -{\textrm{det}}(Rn-R_{2})\nonumber \\
 & = & -{\textrm{det}}(R)(n-n_{+})(n-n_{-}).
\end{eqnarray}
 The integral in (\ref{eq:p0dicho_gen}) can be calculated. The explicit solution
is 
\begin{equation}
\label{p0ndicho}
p_{0}(n)=C(n-\tilde{n})(n-n_{-})^{\alpha _{-}-1}(n_{+}-n)^{\alpha _{+}-1}
\end{equation}
 where 
\begin{equation}
\alpha _{-}=\frac{\lambda _{1}r_{00}(n_{0}-n_{-})}{r_{+}r_{-}(n_{+}-n_{-})},
\end{equation}

\begin{equation}
\alpha _{+}=\frac{\lambda _{1}r_{00}(n_{+}-n_{0})}{r_{+}r_{-}(n_{+}-n_{-})},
\end{equation}

\begin{equation}
\tilde{n}=\frac{r_{210}}{r_{10}}=\frac{r_{2+}-r_{2-}}{r_{+}-r_{-}}.
\end{equation}
 One can show that either \( \tilde{n}<n_{-} \) or \( \tilde{n}>n_{+} \),
so that \( p_{0}(n) \) is non-negative for \( n\in [n_{-},n_{+}] \) as it
should be. For \( \lambda _{1}\rightarrow \infty  \) one obtains the expected
result \( p_{0}(n)=\delta (n-n_{0}) \) where \( n_{0}=r_{200}/r_{00} \). For
small \( \lambda _{1} \), \( p_{0}(n) \) has algebraic singularities at the
boundaries \( n_{\pm } \). For small but finite \( \lambda _{1} \), the singularities
remain integrable. In the limit \( \lambda _{1}\rightarrow 0 \), \( p_{0}(n) \)
tends to a sum of two \( \delta  \)-functions (with appropriate weights according
to the dichotomous process), located at \( n_{\pm }=r_{2\pm }/r_{\pm } \),
\begin{equation}
p_{0}(n)=p_{+}\delta (n-n_{+})+p_{-}\delta (n-n_{-}).
\end{equation}
 The average value \( \bar{n} \) of \( n \) can be calculated explicitely.
One obtains 
\begin{eqnarray}
\bar{n} & = & \int _{n_{-}}^{n_{+}}np_{0}(n)dn\nonumber \\
 & = & \frac{\int _{n_{-}}^{n_{+}}n(n-\tilde{n})(n-n_{-})^{\alpha _{-}-1}(n_{+}-n)^{\alpha _{+}-1}dn}{\int _{n_{-}}^{n_{+}}(n-\tilde{n})(n-n_{-})^{\alpha _{-}-1}(n_{+}-n)^{\alpha _{+}-1}dn}\nonumber \\
 & = & \frac{n_{-}(n_{-}-\tilde{n})B(\alpha _{+},\alpha _{-})+(2n_{-}-\tilde{n})(n_{+}-n_{-})B(\alpha _{+},\alpha _{-}+1)+(n_{+}-n_{-})^{2}B(\alpha _{+},\alpha _{-}+2)}{(n_{-}-\tilde{n})B(\alpha _{+},\alpha _{-})+(n_{+}-n_{-})B(\alpha _{+},\alpha _{-}+1)}\nonumber \\
 & = & \frac{n_{-}(n_{-}-\tilde{n})+(2n_{-}-\tilde{n})(n_{0}-n_{-})+(n_{+}-n_{-})(n_{0}-n_{-})\frac{(\alpha _{-}+1)}{(\alpha _{+}+\alpha _{-}+1)}}{n_{0}-\tilde{n}}.
\end{eqnarray}
 One can easily show that \( \bar{n}=n_{0} \) in the limit \( \lambda _{1}\rightarrow \infty  \),
and \( \bar{n}=p_{+}n_{+}+p_{-}n_{-} \) for \( \lambda _{1}=0 \). Finally
one obtains 
\begin{equation}
\label{nbar_dicho}
\bar{n}=n_{0}+(p_{+}n_{+}+p_{-}n_{-}-n_{0})\frac{\tau }{\bar{\tau }+\tau },
\end{equation}
 where \( \tau =\lambda _{1}^{-1} \) and \( \bar{\tau }=\frac{r_{00}}{r_{+}r_{-}}=\frac{p_{+}}{r_{-}}+\frac{p_{-}}{r_{+}} \).
Thus, \( \bar{n} \) depends monotonously on \( \tau  \). The expression (\ref{nbar_dicho})
as well as the formula for \( p_{0}(n), \) (\ref{p0ndicho}) have been mentioned
without a detailed derivation in \cite{Mielke99}.

\subsection{The general two-state model}

For a general noise process \( z(t) \) the effective rates are 
\begin{equation}
\bar{r}_{i}=\int dz(1-r(z)M_{z}^{-1})^{-1}r_{i}(z)\phi _{0}(z),
\end{equation}
 where again \( r_{1}:=r_{21}=-r_{11} \), \( r_{2}:=r_{12}=-r_{22} \). This
yields directly 
\begin{equation}
\label{eq:naverage}
\bar{n}=\frac{\int dz(1-r(z)M_{z}^{-1})^{-1}r_{2}(z)\phi _{0}(z)}{\int dz(1-r(z)M_{z}^{-1})^{-1}r(z)\phi _{0}(z)}.
\end{equation}
 Whether or not this expression can be calculated explicitely depends on the
generator \( M_{z} \) of the noise process \( z(t) \). A case where the calculation
is possible is a kangaroo process. A kangaroo process is a process where all
non-vanishing eigenvalues of \( M_{z} \) are equal. In that case (\ref{eq:naverage})
simplifies to 
\begin{equation}
\label{eq:navKangaroo}
\bar{n}=\frac{\int dz(1+\tau r(z))^{-1}r_{2}(z)\phi _{0}(z)}{\int dz(1+\tau r(z))^{-1}r(z)\phi _{0}(z)}.
\end{equation}
 The derivation of the last result is not straight forward: One cannot simply
replace \( M_{z} \) by \( -\tau ^{-1} \). But it turns out, that the result
is the same, since additional factors in the denominator and in the numerator
cancel each other. (\ref{eq:navKangaroo}) is valid for a dichotomous process
as well and yields directly (\ref{nbar_dicho}).

\subsection{Comparison with the Fokker-Planck equation}

As already mentioned, the validity of rate equations for fluctuating potentials
is less clear than for fixed potentials. It is therefore important to compare
results for rate equations with exact solutions of the Fokker-Planck equation.
They can be obtained easily for one-dimensional, piecewise linear potentials.
This has been shown in detail for periodic potentials in \cite{Mielke95a, Mielke95b}.
In the present case the procedure is similar, let us therefore describe it only
briefly. For this comparison, we restrict ourselves to the stationary distribution
\( p_{0}(x)=\int dz\rho (x,z) \) and to the mean escape times for the particle
sitting in a minimum of the potential. 

We first divide the domain \( \Omega  \), on which the Fokker-Planck equation
is defined, into a set of intervals \( I_{i} \), \( i=1,\ldots ,N_{I} \).
We assume that the force \( f(x,z)=-\frac{\partial V(x,z)}{\partial x} \) is
constant on each interval \( I_{i} \),
\begin{equation}
f(x,z)=f_{i}(z)\quad \mbox {if}\, x\in I_{i}.
\end{equation}
 As a second step, we expand the stationary solution of the Fokker-Planck equation
using the eigenbasis of \( M_{z} \),
\begin{equation}
\rho (x,z)=p_{0}(x)\phi _{0}(z)+\sum _{k>0}p_{k}'(x)\phi _{k}(z).
\end{equation}
 On the interval \( I_{i} \), the Fokker-Planck equation yields
\begin{eqnarray}
f_{i}(z)p_{0}'(x)\phi _{0}(z)+Tp_{0}''(x)\phi _{0}(z)+\sum _{k>0}p_{k}''(x)f_{i}(z)\phi _{k}(z) &  & \nonumber \\
+T\sum _{k>0}p_{k}'''(x)\phi _{k}(z)-\sum _{k>0}\lambda _{k}p_{k}'(x)\phi _{k}(z) & = & 0.
\end{eqnarray}
 This equation can be integrated once with respect to \( x \). Furthermore,
expression of the form \( f_{i}(z)\phi _{k}(z) \) can be expanded,
\begin{equation}
f_{i}(z)\phi _{k}(z)=\sum _{k'}f^{(i)}_{k,k'}\phi _{k'}(z).
\end{equation}
 One then obtains
\begin{equation}
\label{rec1}
f^{(i)}_{0,0}p_{0}(x)+\sum _{k'>0}f^{(i)}_{k',0}p_{k'}'(x)+Tp_{0}'(x)=0,
\end{equation}
 
\begin{equation}
\label{rec2}
f^{(i)}_{0k}p_{0}(x)+\sum _{k'>0}f_{k',k}^{(i)}p_{k'}'(x)+Tp_{k}''(x)=\lambda _{k}p_{k}(x),\quad k>0.
\end{equation}
 The coefficients \( f^{(i)}_{k,k'} \) and therefore the final set of equations
for \( p_{k}(x) \) do not depend on the representation of the noise process
\( z(t) \) as ist should be. If the space of functions \( \phi _{k}(z) \)
has finite dimension \( N \), (\ref{rec1}) and (\ref{rec2}) is a set of \( N \)
differential equation with constant coefficients, which can be solved explicitely.
One obtains a simple eigenvalue problem for an \( N\times N \) matrix, which
can be solved numerically. A general solution of this problem is a linear combination
of the \( N \) different solutions of the eigenvalue problem. The remaining
problem is thus to determine the \( N_{I}N \) coefficients in these linear
combinations. They are determined by the normalization of \( p_{0}(x) \), by
the continuity of \( p_{0}(x) \) and by the continuity of \( p_{k}(x) \) and
\( p_{k}'(x) \) for \( k>0 \). This is a problem of solving \( N_{I}N \)
linear equations for \( N_{I}N \) unknown coefficients, which can again be
done easily numerically as long as \( N_{I}N \) is not too large. 
\begin{figure}
\resizebox*{0.7\textwidth}{0.35\textheight}{\rotatebox{270}{\includegraphics{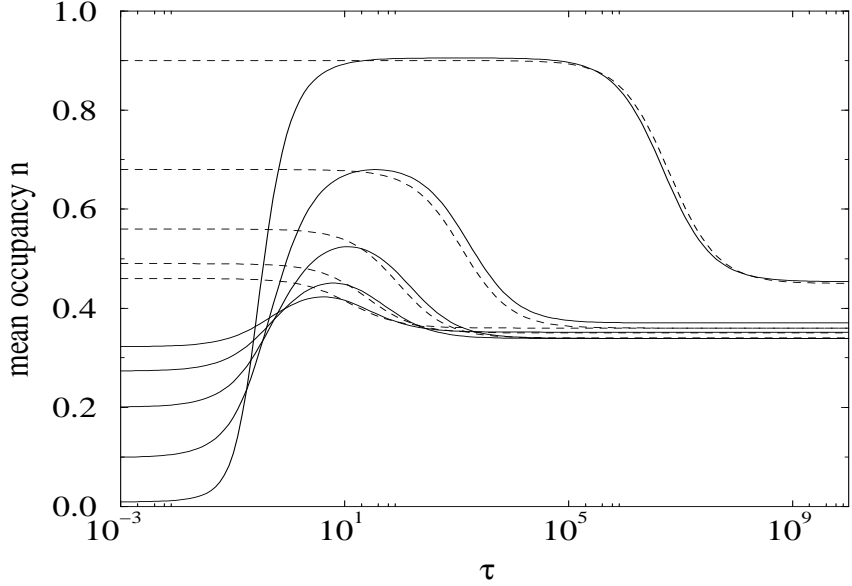}}}

\caption{The mean occupancy in the left minimum of a piecewise linear, dichotomously
fluctuating potential as a function of \protect\( \tau \protect \) for various
\protect\( T\protect \) . The parameters are \protect\( T=0.2,\, 0.4,\, 0.6,\, 0.8,\, 1.0\protect \),
the average force takes the values 10, 3/2, -3/2, 4/3, -4, -10, \protect\( \Delta f\protect \)
takes the values 0, 1/4, -1/4, 2/3, -2, 0. The force jumps at the \protect\( x=-4,\, -2,\, 0,\, 3,\, 4\protect \).
The maximal value of the mean occupancy decreases with increasing \protect\( T\protect \).
The dashed lines are the results from the effective rate equation.}
\end{figure}
In Fig. 1 we show explicit results for a one-dimensional dichotomous two-state
model. The potential \( V(x,z)=\langle V\rangle (x)+z\Delta V(x) \) has two
minima. \( z(t) \) is a dichotomous process which takes the two values \( \pm 1 \).
In this case \( N=2 \) and \( N_{I}=6 \). The plot shows the probability to
find the particle in the left potential well as a function of \( \tau  \).
It can be compared to \( \bar{n} \), which can be calculated from the effective
rate equations. A similar comparison for a different, dichotomously fluctuating
potential was shown in \cite{Mielke99}. The results from the rate equation
yield an accurate approximation for sufficiently large \( \tau  \), as expected.
The intra-well relaxation time can be estimated from the time, the particle
needs to reach a minimum of the potential if it starts close to the maximum.
In our units it is somewhat larger than unity. The approximation becomes better
for smaller temperature. The behaviour is similar for other dichotomously fluctuating
potentials with two minima.

For other stochastic processes \( z(t) \) one can draw similar conclusions.
Let us consider as an example a potential \( V(x,z)=\langle V\rangle (x)+z\Delta V(x) \)
where now \( z(t) \) is a sum of \( N-1 \) dichotomous processes, where each
dichotomous process takes the values \( \pm 1/\sqrt{N} \). This means that
\( z(t) \) takes the values \( (-N+2n+1)/\sqrt{N} \), \( n=0,\ldots ,N \)
with the probability \( p_{n}=2^{-N}\left( \begin{array}{c}
N\\
n
\end{array}\right)  \). In the limit \( N\rightarrow \infty  \) one obtains an Ornstein-Uhlenbeck
process \cite{Maruyama88}. For \( \langle V\rangle  \) and \( \Delta V \)
we take the same values as in Fig. 1. Results for the mean occupancy are show
in Fig. 2 together with the corresponding results from the rate equation, obtained
from (\ref{eq:naverage}). As for the dichotomous process, the results from
the rate equation agree quite well with the results from the Fokker-Planck equation
for not too small values of \( \tau  \).
\begin{figure}
{\par\centering \resizebox*{0.7\textwidth}{0.35\textheight}{\rotatebox{270}{\includegraphics{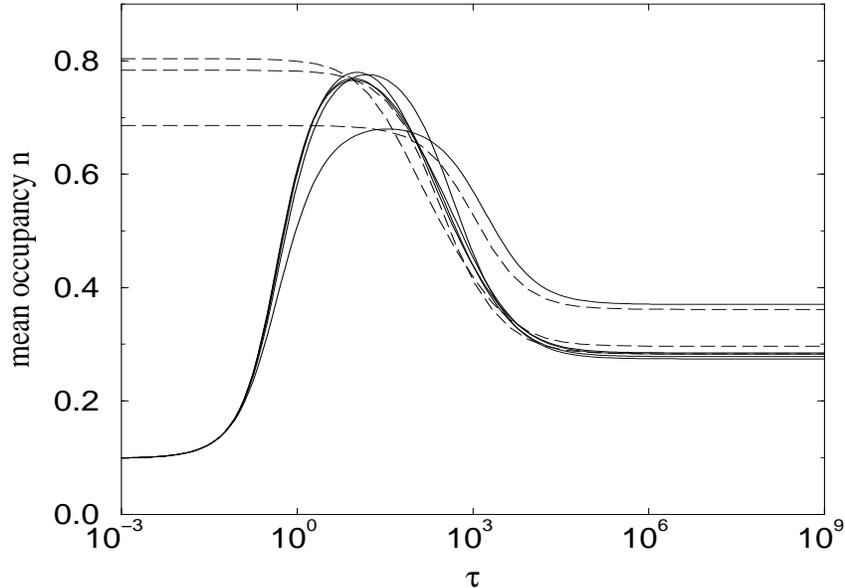}}} \par}

\caption{The mean occupancy in the left minimum for a fluctuating potential as in Fig.
1. \protect\( f\protect \) and \protect\( \Delta f\protect \) are the same
as in Fig. 1, \protect\( T=0.4\protect \). The different curves are for sums
of \protect\( N-1\protect \) dichotomous processes, \protect\( N=2,\ldots ,6\protect \).
For increasing \protect\( N\protect \) the maximum of the curve is shifted
to the left. The dashed lines are the results from the effective rate equation,
the results for \protect\( N=4,5,6\protect \) cannot be distinguished. }
\end{figure}
Similar results can be obtained for other noise processes as well. For kangaroo
processes, the mean occupancy has been obtained explicitely (\ref{eq:navKangaroo}).
The corresponding results for the mean occupancy are similar to what has been
obtained so far. If one takes for instance kangaroo processes with the same
\( \phi _{0}(z) \) as for the sums of dichotomous processes in Fig. 2, the
quantitative results are almost the same, they differ by less than \( 1\% \). 

The results in Fig. 2 show that already for small \( N \) one obtains essentially
the result for the Ornstein-Uhlenbeck process. The convergency to the Ornstein-Uhlenbeck
process is very fast. For \( N>6 \) the curves lie on top of the curve for
\( N=6 \). Clearly, the convergency depends on the parameters of the system.
For smaller temperatures we observe a slower convergency.

It is possible to consider potentials with more than two minima as well. The
corresponding rate equations describe systems with more than two states. For
simple systems with few minima and e.g. a dichotomously fluctuating potential,
one can again compare the results from the Fokker-Planck equation with the results
from the rate equations. The generic behaviour of such systems is again described
by the effective rates as long as \( \tau  \) is not too small. Whether or
not (\ref{eq:effrates}) can be used to calculate the effective rates analytically
depends on the structure of the problem. For complicated noise processes or
for potentials with many minima, \( R_{\rm eff} \) can only be obtained numerically.

\section{Mathematical Aspects}

In the main part of this work we investigate the stationary solution of a Fokker-Planck
equation (\ref{eq:FokkerPlanck}) with a fluctuating potential, assuming that
such a stationary solution exists. The aim of this section is to proof the existence
of a stationary solution under certain assumptions that will be sufficient for
our purpose. For the special case of a one-dimensional, dichotomously fluctuating
potential, Pechukas and Ankerhold \cite{Pechukas98} proved the existence of
a stationary solution under some special assumptions on the potential. The result
of this section can be understood as a generalization of their results.

\subsection{The discrete case}

Let us first discuss the discrete case, which is very simple but shows some
interesting aspects. Let us assume that a system has \( N \) states and that
it can move from one state to another. The behaviour is described by a rate
equation
\begin{equation}
\frac{dp_{i}}{dt}=\sum _{j}r_{ij}p_{j}.
\end{equation}
We now assume that the rates fluctuate between \( N_{r} \) different states
\( \alpha =1,\ldots ,N_{r} \). The fluctuation between these states is determined
by a matrix \( M=(m_{\alpha \beta })_{\alpha ,\beta =1,\ldots ,N_{r}} \) that
contains the rates for the fluctuations. The probability \( p_{i\alpha } \)
of the fluctuating system to be in the state given by \( i \) and \( \alpha  \)
is determined by 
\begin{equation}
\label{eq:rate-general}
\frac{dp_{i\alpha }}{dt}=\sum _{j}r_{ij\alpha }p_{j\alpha }+\sum _{\beta }m_{\alpha \beta }p_{i\beta }.
\end{equation}
 The stationary solution of this equation, if it exists, must be the eigenvector
of the matrix
\begin{equation}
\hat{R}=\left( \begin{array}{ccccc}
R_{1}+m_{11}I & m_{12}I & m_{13}I & \cdots  & m_{1N_{r}}I\\
m_{21}I & R_{2}+m_{22}I & m_{23}I & \cdots  & \\
m_{31}I & m_{32}I & R_{3}+m_{33}I & \ddots  & \vdots \\
\vdots  &  & \ddots  & \ddots  & m_{N_{r}-1,N_{r}}I\\
m_{N_{r}1}I & \cdots  &  & m_{N_{r},N_{r}-1}I & R_{N_{r}}+m_{N_{r}N_{r}}I
\end{array}\right) 
\end{equation}
to eigenvalue 0. Here, \( I \) is the \( N\times N \) unit matrix. The matrices
\( R_{\alpha } \) are rate-matrices. As a consequence, 
\begin{equation}
r_{ii\alpha }=-\sum _{j\neq i}r_{ji\alpha }
\end{equation}
 and \( r_{ij\alpha }\geq 0 \) for \( i\neq j \). Furthermore, \( M \) is
a rate matrix, \( m_{\alpha \alpha }=-\sum _{\beta \neq \alpha }m_{\beta ,\alpha } \)
and \( m_{\alpha ,\beta }\geq 0 \) for \( \alpha \neq \beta  \). Therefore,
all the off-diagonal matrix elements of \( \hat{R} \) are non-negative. I assume
that starting from some state of the system, any other state can be reached
dynamically. This means that \( \hat{R} \) is irreducible. Then, one can introduce
a constant \( c \) that is larger than the modulus of any diagonal matrix element
of \( \hat{R} \). Let \( \hat{I} \) be the \( N_{r}N\times N_{r}N \) unit
matrix. Then the matrix \( \hat{R}+c\hat{I} \) is irreducible and has only
positive matrix elements. As a consequence, the Perron-Frobenius theorem applies.
The eigenvalue with the largest modulus of \( \hat{R}+c\hat{I} \) is non-degenerate,
real, and positive, and the corresponding left and right eigenvectors have only
non-negative entries. Now, the vector \( (1,1,\ldots ,1) \) is a left eigenvector
of \( \hat{R} \) with the eigenvalue \( c \). Therefore \( c \) is the eigenvalue
with the largest modulus of \( \hat{R}+c\hat{I} \). This shows, that \( 0 \)
is the eigenvalue of \( \hat{R} \) with the largest real part and that its
right eigenvalue has only non-negative entries. It is the stationary solution
of (\ref{eq:rate-general}). One can show that any solution of the rate equation
(\ref{eq:rate-general}) tends to the stationary solution for \( t\rightarrow \infty  \). 

This proof is very simple, it is based on the fact that rates are non-negative
and on the Perron-Frobenius theorem. Let us now take a look at the original
Fokker-Planck equation (\ref{eq:FokkerPlanck}). For simplicity, we discuss
only the case where the potential fluctuates between \( N_{r} \) different
states. \( z \) takes \( N_{r} \) different values. Let us assume that the
Fokker-Planck equation is defined on a finite open domain \( \Omega  \) with
reflecting boundary conditions on the boundary \( \partial \Omega  \). If one
introduces some kind of coarse-graining, i.e. a partition of \( \Omega  \)
into \( N \) small parts \( \Omega _{i} \), it is then possible to apply the
above result and a stationary solution exists. This holds for any partition
of \( \Omega  \) into small parts. Therefore one should expect that a stationary
solution of the Fokker-Planck equation (\ref{eq:FokkerPlanck}) exists as well.
We will show this in the next subsection using a different approach. One has
to show that the Fokker-Planck operator has a unique right eigenstate with eigenvalue
\( 0 \) and that any solution of the Fokker-Planck equations tends to that
solution for \( t\rightarrow \infty  \). Since we are dealing with a differential
operator and not with a matrix, it is not possible to use a Perron-Frobenius
type argument.

\subsection{The continuous case}

Let us first show that the ratio of any two solutions of the Fokker-Planck equation
tends to unity in the limit \( t\rightarrow \infty  \). As for the discrete
case we assume that \( z \) takes \( N_{r} \) values \( z_{\alpha } \), \( \alpha =1,\ldots ,N_{r} \)
and we let
\begin{equation}
\rho (\vec{x},z_{\alpha },t)=p_{\alpha }(\vec{x},t).
\end{equation}
Furthermore, we again assume that the Fokker-Planck equation (\ref{eq:FokkerPlanck})
is defined on a finite open domain \( \Omega  \) with reflecting boundary conditions
on the boundary \( \partial \Omega  \). The Fokker-Planck equation can be written
in the form
\begin{equation}
\frac{\partial p_{\alpha }}{\partial t}=\nabla (\nabla V_{\alpha }(\vec{x})+T\nabla )p_{\alpha }+\sum _{\beta }m_{\alpha \beta }p_{\beta }=L_{\alpha }p_{\alpha }+\sum _{\beta }m_{\alpha \beta }p_{\beta }.
\end{equation}
 Let \( p_{1\alpha } \), \( p_{2\alpha } \) be two positive solutions of this
equation. We define 
\begin{equation}
H(t)=\sum _{\alpha }\int d^{d}x\, p_{1\alpha }\ln (p_{1\alpha }/p_{2\alpha }).
\end{equation}
 This function is often called Kullback information and has been introduced
by Kullback \cite{Kullback51, Kullback}. By standard methods one can show that
(i) \( H(t)\geq 0 \) and (ii) \( \frac{dH}{dt}\leq 0 \). As a consequence
\( H \) tends to zero for \( t\rightarrow \infty  \) and therefore \( p_{1\alpha }/p_{2\alpha } \)
tends to \( 1 \). The proof is similar to the one presented in chapter 6 of
\cite{Risken}. First, one shows 
\begin{equation}
H(t)=\sum _{\alpha }\int d^{d}x\, p_{2\alpha }\left( \frac{p_{1\alpha }}{p_{2\alpha }}\ln \frac{p_{1\alpha }}{p_{2\alpha }}-\frac{p_{1\alpha }}{p_{2\alpha }}+1\right) \geq 0,
\end{equation}
 which follows from \( r\ln r-r+1\geq 0 \) for \( r\geq 0 \). Now, one can
calculate
\begin{eqnarray}
\frac{dH}{dt} & = & \sum _{\alpha }\int d^{d}x(\frac{dp_{1\alpha }}{dt}\ln (p_{1\alpha }/p_{2\alpha })-(p_{1\alpha }/p_{2\alpha })\frac{dp_{2\alpha }}{dt})\nonumber \\
 & = & \sum _{\alpha }\int d^{d}x\left( (L_{\alpha }p_{1\alpha })\ln (p_{1\alpha }/p_{2\alpha })+\sum _{\beta }m_{\alpha \beta }p_{\beta }\ln (p_{1\alpha }/p_{2\alpha })-\frac{p_{1\alpha }}{p_{2\alpha }}\frac{dp_{2\alpha }}{dt}\right) \nonumber \\
 & = & \sum _{\alpha }\int d^{d}x\left( p_{1\alpha }(L^{\dagg }_{\alpha }\ln (p_{1\alpha }/p_{2\alpha })+\sum _{\beta }m_{\alpha \beta }p_{\beta }\ln (p_{1\alpha }/p_{2\alpha })-\frac{p_{1\alpha }}{p_{2\alpha }}\frac{dp_{2\alpha }}{dt}\right) .
\end{eqnarray}
 The first term may be evaluated using
\begin{equation}
L_{\alpha }^{\dagg }\ln (p_{1\alpha }/p_{2\alpha })=\frac{p_{2\alpha }}{p_{1\alpha }}L_{\alpha }^{\dagg }(p_{1\alpha }/p_{2\alpha })-T\frac{p^{2}_{2\alpha }}{p^{2}_{1\alpha }}(\nabla (p_{1\alpha }/p_{2\alpha }))^{2}.
\end{equation}
 This yields
\begin{equation}
\frac{dH}{dt}=-T\sum _{\alpha }\int d^{d}x\, p_{1\alpha }(\nabla \ln (p_{1\alpha }/p_{2\alpha }))^{2}+\int d^{d}x\sum _{\alpha \beta }m_{\alpha \beta }\left( p_{1\beta }\ln \frac{p_{1\alpha }}{p_{2\alpha }}-p_{2\beta }\frac{p_{1\alpha }}{p_{2\alpha }}\right) .
\end{equation}
 The first term is clearly negative or \( 0 \). To investigate the second term,
let us introduce
\begin{equation}
\Lambda _{\alpha \beta }=\delta _{\alpha \beta }+\epsilon m_{\alpha \beta },
\end{equation}
 where \( \epsilon  \) should be sufficiently small (\( \epsilon ^{-1}>\max _{\alpha }|m_{\alpha \alpha }| \),
I will take the limit \( \epsilon \rightarrow 0 \) below). Let \( \hat{p}_{i\alpha }=\sum _{\beta }\Lambda _{\alpha \beta }p_{i\beta } \).
One can show that
\begin{eqnarray}
\int d^{d}x\sum _{\alpha }\left( p_{1\alpha }\ln \frac{p_{1\alpha }}{p_{2\alpha }}-\hat{p}_{1\alpha }\ln \frac{\hat{p}_{1\alpha }}{\hat{p}_{2\alpha }}\right)  & = & \int d^{d}x\sum _{\alpha \beta }\Lambda _{\alpha \beta }p_{1\beta }\ln \frac{p_{1\beta }\hat{p}_{2\alpha }}{p_{2\beta }\hat{p}_{1\alpha }}\nonumber \\
 & \geq  & \int d^{d}x\sum _{\alpha \beta }\Lambda _{\alpha \beta }p_{1\beta }\left( 1-\frac{p_{2\beta }\hat{p}_{1\alpha }}{p_{1\beta }\hat{p}_{2\alpha }}\right) \, =\, 0.
\end{eqnarray}
 Expanding the left hand side in powers of \( \epsilon  \), one obtains 
\begin{equation}
\int d^{d}x\sum _{\alpha \beta }m_{\alpha \beta }\left( p_{1\beta }\ln \frac{p_{1\alpha }}{p_{2\alpha }}-p_{2\beta }\frac{p_{1\alpha }}{p_{2\alpha }}\right) \leq 0
\end{equation}
 and therefore \( \frac{dH}{dt}\leq 0 \). The limit \( p_{\alpha }(\vec{x},t\rightarrow \infty ) \)of
a positive solution of the Fokker-Planck equation is positive. If the potential
\( V_{\alpha }(\vec{x}) \) is bounded within the domain \( \Omega  \), and
if \( \Omega  \) is connected, this limit yields the unique stationary solution
of the Fokker-Planck equation, since the Fokker-Planck operator does not depend
on time. This shows the existence of the stationary solution of the Fokker-Planck
equation.

\section{Outlook and conclusions}

One main result of this paper is that the over-damped motion of a Brownian particle
in a fluctuating potential can be described by kinetic rate equations if (i)
the potential fluctuations are slower than the relaxation of the particle within
a minimum of the potential and if (ii) the positions of the minima of the potential
do not fluctuate. For temperatures small compared to typical barrier heights
of the potential, the quantitative agreement of the two different descriptions
is very good if one calculates stationary or quasi-stationary properties. This
shows that the long-time behaviour of a Brownian particle in a fluctuating potential
has universal properties and does not depend on the details of the potential
but only one the rates for the transition over the various fluctuating barriers.
This is important, since in many realistic situations details of the potential
are not known. For instance, in the case of a cell surface receptor or some
other protein in a cell membrane, one knows eventually something about the stable
or metastable conformations of the protein, but (almost) nothing about the potential
that describes the energy of the deformations of the protein. Even if the potential
is known, it is often simpler to solve a kinetic rate equation instead of a
Fokker-Planck equation. 

Starting from the kinetic rate equation with fluctuating rates, we derived a
formula for effective rates which holds for general noise processes. This formula
yields directly information on the stationary and long-time properties of the
system. It can be evaluated for many noise processes. The special case of a
dichotomous two-state system can be solved exactly and one obtains the distribution
functions for the occupancy of the two states.

As a byproduct, we could show the existence of a stationary solution of the
Fokker-Planck equation with a fluctuating potential under some general conditions
and thereby generalize a result published recently by Pechukas and Ankerhold
\cite{Pechukas98}. 

As a direct consequence of our calculations, one can investigate where rate
equations can yield good results and where not. This is important since in many
applications, mainly in biological situations, rate equations are used for a
direct comparison with experiments. For instance, dynamical properties of membrane
proteins are often described and investigated by rate equations. An interesting
and well known example is the effect of time dependent electric fields on such
proteins. It has been discussed it detail by Astumian and Robertson \cite{Astumian89}
using rate equations. In such systems one observes strong amplifications of
weak signals. Moss \cite{Moss91} suggested that the reason may be stochastic
resonance. Kruglikov and Dertinger \cite{Kruglikov94} gave a qualitative discussion
supporting the occurence of stochastic resonance in such systems using a time-dependent
potential, but a description using rate equations would have been possible as
well. As already mentioned, Petracchi et al \cite{Petracchi94} studied the
effect of time dependent electric fields on membrane proteins experimentally.
As a specific example they used a K\( ^{+} \) channel. They measured various
stationary probabilities and dynamical quantities to describe the statistics
of the transitions in these systems. As an interesting result they found a phase
anticipation: One of the transitions occurs with a negative phase shift compared
to the stimulus. They compared their experimental results with numerical results
for a simple two-state Markov model, described by a rate equation. The rate
equations do not show the phase anticipation. The authors argued that this effect
has a biological origin and discuss various possible hypotheses to explain it.

From our discussion it is clear that rate equations do describe the stationary
or long-time behaviour of a system quite well. Indeed, the stationary or quasi-stationary
properties calculated by Petracchi et al \cite{Petracchi94} agree very well
with their experimental findings. The phase anticipation is a dynamical effect
that occurs on shorter time scales. Therefore one cannot expect that it can
be obtained using rate equations. Our calculations suggest that this effect
depends on the details of the potential and not only on the rates, which are
determined mainly by the barrier heights. The main problem is that in the case
of a specific membrane protein the potential that describes the dynamics is
not known. Therefore it is not possible to model the K\( ^{+} \) channel using
a time-dependent potential. But it would be of general interest to investigate
whether or not and under which conditions a Brownian particle in a time-dependent
potential shows the described phase anticipation. This is clearly beyond the
scope of the present paper.

\subsection*{Acknowledgement}

I wish to thank H. Dertinger for various interesting discussions on the effect
of weak electromagnetic fields on cells.

\end{document}